# Cooperative Intergroup Mating Can Overcome Ethnocentrism in Diverse Populations

Caitlin J. Mouri*
mouri.caitlin@gmail.com

Thomas R. Shultz*^
thomas.shultz@mcgill.ca

*Department of Psychology, McGill University
^School of Computer Science, McGill University

## Abstract

Ethnocentrism is a behavioral strategy seen on every scale of social interaction. Game-theory models demonstrate that evolution selects ethnocentrism because it boosts cooperation, which increases reproductive fitness. However, some believe that interethnic unions have the potential to foster universal cooperation and overcome in-group biases in humans. Here, we use agent-based computer simulations to test this hypothesis. Cooperative intergroup mating does lend an advantage to a universal cooperation strategy when the cost/benefit ratio of cooperation is low and local population diversity is high.

**Key Words:** ethnocentrism, cooperation, evolution, diversity.

## Introduction

Ethnocentrism is a behavioral strategy in which individuals cooperate with members of their own group, but not with outsiders [1]. In-group favoritism is seen on every scale of social interaction, from the geopolitical [2] to the interpersonal [1]. In humans, its consequences are severe: ethnic clashes characterize 75% of armed conflicts since the end of the Cold War [2], highlighting the importance of addressing ethnocentric tendencies.

Empirical observations suggest that ethnocentrism is a fundamental cooperative strategy. In-group preference in humans can be activated through arbitrary group tags, with little social or biological relevance [3, 4]. Nor do such biases require complex cognitive capacity, as they can be observed among microbes [5], plants [6], and insects [7].

Game theoretic research on symmetric games such as prisoner's dilemma (PD) shows that mutual defection is an evolutionary stable strategy in well-mixed populations, because a player can achieve better fitness payoffs by defecting, regardless of what the other player does [8]. Empirical and theoretical work shows that cooperation can evolve by either direct [9-11] or indirect reciprocity (based on reputation) [12, 13], but only if the game is indefinitely repeated. Evolution of cooperation in one-off PD games can occur via some kind of group structure, based on either spatial connections [14, 15] or tags that probabilistically identify agent kinship or similarity [16, 17]. Multi-level selection can also favor groups that cooperate within the group and compete across groups [18, 19]. In evolutionary simulations with identifiable group tags, ethnocentric agents out-compete humanitarians (universal cooperators) by defecting against altruistic members of other groups [16, 17]. This strengthens both within-group cooperation and between-group competition, providing ethnocentrism with a powerful evolutionary impetus. Evolutionary simulations also show that diminishing in-group biases hinder cooperative behavior [20].

However, models that rely on evolutionary game theory often ignore some relevant real-world behaviors [21]. Mate selection is one such behavior. Humans show a tendency towards ethnocentric mate selection, choosing mates who are similar to themselves genetically [22, 23], religiously [24], and in personality [25]. In turn, assortative mating, based on similarity, may strengthen ethnocentrism by accentuating genetic differences [26-28], decreasing mobility between groups [29], and sustaining cultural differentiation [30]. The causal relationship between ethnocentric mate selection and more general ethnocentric behavior has not been explored. Nor can it be taken for granted that assortative mating is simply a byproduct of general ethnocentric tendencies. In fact, primate researchers have recently suggested a reverse causality, whereby the evolutionary advantages of cooperative breeding may lead to the evolution of general cooperative behaviors [31]. This suggests that the fitness of cooperative behaviors is directly influenced by their success in mating contexts. While ethnocentrism has been thoroughly tested for evolutionary fitness in non-mating contexts, it has not, to our knowledge, been tested in a mating context. This would add a layer of complexity to the evolution of ethnocentrism, by introducing the potential for intergroup mating and multi-group identities. Some researchers have speculated that these factors could undermine ethnocentrism by introducing diversity into local kinship networks [32, 33].

By extending agent-based computer simulations that have previously been used to study ethnocentric behavior [16, 17], we test the evolutionary fitness of ethnocentric mate selection. Rather than reproducing agents via the usual technique of cloning, agents in our simulations mate and exchange genetic material. We explore intergroup mating and offspring identities for their potential effects on the evolution of cooperative behavior, with some emphasis on the relative success of ethnocentric and humanitarian strategies. We discuss our results in light of current ideas about interethnic mating in humans.

# Methods

We extend agent-based computer simulations [16, 17] to accommodate mating. The simulation begins with an empty 50x50 toroid lattice. Forty-eight agents with the same group tag are placed into random locations within each quadrant of the lattice, such that each quadrant initially contains agents from one group only. These initial agents are randomly assigned one of four cooperative strategies. Agents who cooperate with agents in their in-group and their out-group are called *humanitarian*. Agents who cooperate only with agents in their in-group, but not agents in their out-group, are termed *ethnocentric*. Agents who cooperate only with agents in their out-group are labeled *traitorous*. Finally, agents who cooperate with no other agents are designated *selfish*. The simulation iterates through evolutionary cycles in which agents interact, reproduce, and die. One thousand evolutionary cycles are run for each simulation, by which time population genotypes have stabilized.

During the interaction phase, agents engage in a one-off prisoner's dilemma game with each of their Moore neighbors (Figure 1). The outcome of each game for each agent is the benefit of receiving cooperation minus the cost of giving cooperation (see Table 1). This net outcome is added to each agent's reproductive potential. To vary the critical cost/benefit ratio, we fix benefit at .1 and vary cost from .01 to .1 in ten steps of .01.

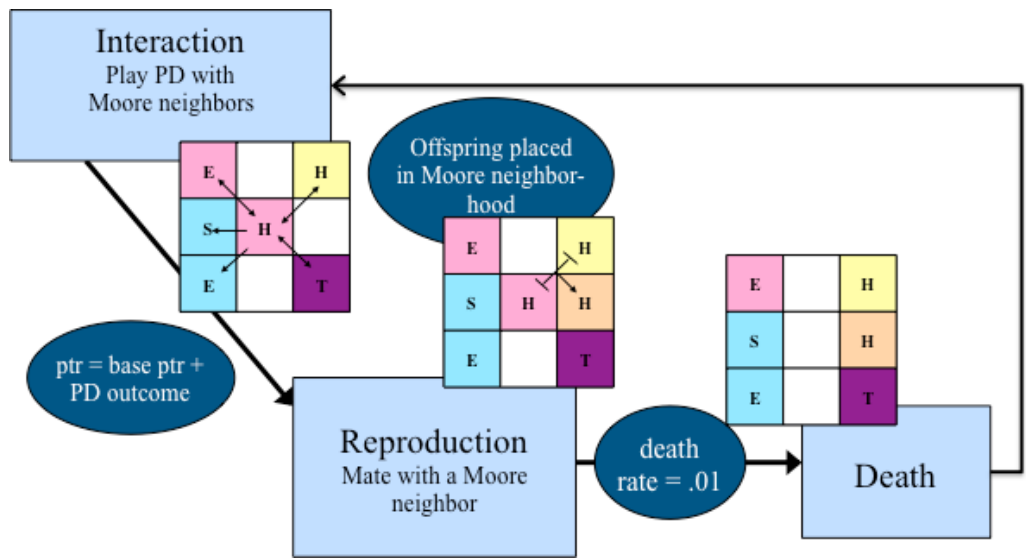

Figure 1. At each evolutionary cycle, each agent plays a PD game with each of its adjacent neighbors. The agent plays according to its inherited cooperative strategy (humanitarian, ethnocentric, traitorous, or selfish). The outcome of the game slightly modifies the agent's reproductive potential. In the next phase, agents have an opportunity to reproduce. An agent mates with a neighboring agent if at least one of them is a willing cooperator. In the Figure, direction of cooperation is indicated by small arrows, and group tags are represented by background color. Parents produce an offspring with a probability corresponding to their mean reproductive potential. Offspring are placed in a neighboring empty cell. Agents die with a probability corresponding to the death rate. Dead agents are removed.

During the reproduction phase, agents are paired randomly with one of their Moore neighbors. If at least one agent is a willing cooperator, the pair mates with a probability equal to their average reproductive potential. Requiring mutual cooperation for mating produces qualitatively similar results, the only exception being that selfishness quickly goes to extinction. Offspring are placed in an empty location in the parents' Moore neighborhood.

Table 1. Outcomes, in terms of cost and benefit, for Agent 1 in a PD game between Agent 1 and Agent 2. The outcome is added to Agent 1's reproductive potential.

|  | Agent 2 | |
|---|---|---|
| Agent 1 | Cooperate | Defect |
| Cooperate | b - c | -c |
| Defect | b | 0 |

Offspring agents' cooperation traits and group tags are inherited from their parent agents. In-group and out-group cooperation traits are each inherited from one of the two parents, randomly selected. We tested three different classification schemes for multiethnic agents, based on human classification of mixed-race individuals [34, 35]. In the *Either/Or* condition, mixed offspring inherit one identity from one or the other randomly-selected parent. In the *Both* and *Neither* conditions, parent agents' tags are averaged together to create a multiethnic tag. For example, if a Group 1 agent is labeled with a vector [1 0 0 0] and a Group 2 agent is labeled with a vector [0 1 0 0], their offspring would have a tag of [.5 .5 0 0]. In the next generation, an agent with a tag of [.5 .5 0 0] could mate with an agent with a tag of [0 1 0 0] to produce an offspring with a tag of [.25 .75 0 0].

The Neither and Both conditions differ in the way these multiethnic agents are grouped. In the Neither condition, the agents must share the same ancestry to be considered in the same group. In the Both condition, agents need to share only some of the same ancestry to be considered in the same group (Figure 2). If less than one quarter of an agents' ancestry originates from a particular group, that portion of its ancestry is ignored during the classification process.

We do not adhere to binary Mendelian genetics in the Both and Neither conditions because ethnicity is not a Mendelian trait that can be traced to a single gene locus. Rather, ethnicity is an amalgam of genetically and culturally transmitted markers. Thus, blended inheritance is an appropriate, if simplistic, model of the way ethnicity can be inherited in humans.

At the end of each reproductive phase, all reproductive potentials are reset to a base of .12. At the end of each cycle, agents die at a rate of .01. Results are obtained by averaging each dependent measure (strategy type, behaviors, and intergroup borders) across the last 100 cycles in each of ten independent simulations.

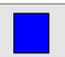

Figure 2. Sample interactions for each offspring-identification scheme. In the Both and Neither conditions, multiethnic offspring are generated when agents with different tags mate with one another. The parents' tags are averaged together, as described in the text. In the Both condition, these offspring are classified with any other agents having Group 1 or Group 2 ancestry. In the Neither condition, the offspring are classified only with agents having Group 1 and Group 2 ancestry. In the Either condition, agents inherit only one tag from one random parent, precluding the existence of multiethnic agents. Check marks indicate cooperation; x indicates defection.

## Results

Each of three dependent measures (strategy proportions of the population, proportion of cooperative behaviors, and number of intergroup borders) is subjected to a repeated-measures ANOVA with offspring ID and cost/benefit ratio as independent factors. For ANOVA of strategies, the four genotypic strategies serve as a repeated measure. The ten replications play the statistical role of subjects in strategy ANOVAs.

In the ANOVA of strategy proportions, between-replication effects include small main effects of ID and c/b ratio, but no interaction. All of the within-replication effects are large (see Table 2 for details). A partial eta squared ($\eta^2$) value above .26 is conventionally considered to indicate a large effect.

Means and standard errors for these effects are presented in Figure 3, plotted separately for the Offspring IDs of Either (3A), Neither (3B), and Both (3C). As a general heuristic, non-overlapping SE bars around each of two means indicate that these two means differ significantly from each other by various multiple comparison procedures. Thus in this paper, standard error bars provide a useful graphical indicator of statistical significance for the difference between any two means in a figure.

For IDs of Either and Neither, c/b ratios of .4 and higher yield the familiar ethnocentric dominance. But with lower c/b ratios, there is clear humanitarian dominance. As shown in Figure 3C, the Both ID condition yields a pattern of consistent ethnocentric dominance except at the highest two c/b values, where selfish genotypes are most prevalent. Interestingly, intergroup borders nearly disappear in the Both condition, as illustrated in Figure 4. All factors in the factorial ANOVA of intergroup borders are significant at p < .001 (see Table 3 for details). In essence, the Both ID scheme eventually creates a single homogenous group, as in the melting pot metaphor. The role of homogeneity in fostering ethnocentrism is discussed later in this section.

Table 2. ANOVA Effects for Strategy Proportions

| Factor | F | df | p < | $\eta^2$ |
|---|---|---|---|---|
| ID | 4.09 | 2, 270 | .05 | .029 |
| c/b | 2.52 | 9, 270 | .01 | .077 |
| ID x c/b | 1.08 | 18, 270 | .368 | .067 |
| Strategy | 3428 | 3, 810 | .001 | .927 |
| Strategy x ID | 61 | 6, 810 | .001 | .312 |
| Strategy x c/b | 237 | 27, 810 | .001 | .888 |
| Strategy x ID x c/b | 33 | 54, 810 | .001 | .690 |

The factorial ANOVA of proportion of cooperative behaviors shows large effects at p < .001 for ID and cost/benefit, as well as their interaction (see Table 4 for details). Means and SE bars are plotted in Figure 5, where it can be seen that cooperation drops off with increasing c/b ratio but never falls below .5. At low levels of c/b ratio, cooperation is quite high.

In order to draw a close comparison to a previous simulation which found that cooperation decreased under humanitarian dominance [20], we focus on two levels of c/b, one of which yields substantial humanitarian dominance (.2) while the other yields substantial ethnocentric dominance (.6) (see Figures 3A and 3B). The relevant means and SE bars are plotted in Figure 6 and the ANOVA details are shown in Table 5. At a c/b ratio of .2, there is strong humanitarian dominance and a higher level of cooperation as compared to the strong ethnocentric dominance and a lower level of cooperation at a c/b ratio of .6. This is a strong difference in the opposite direction from the previous simulation that used increased cognitive cost for conditional strategies like ethnocentrism [20]. Our cooperative intergroup mating technique produces humanitarian dominance and even higher levels of cooperation than is evident when ethnocentrics are dominant.

The general similarity between the Neither and Either ID conditions across strategies (Figures 3A and 3B) and behaviors (Figures 5 and 6) indicates that the presence of agents with multiethnic identities has no effect on genotypic strategies or the amount of cooperation that they produce. Whereas agents are able to form unique, multiethnic identities in the Neither condition, no multiethnic identities exist in the Either condition. This lack of difference between these two conditions in our simulations contradicts claims from correlational evidence with humans that the *one-drop* rule, which retains strict Either/Or identities for multiethnic individuals [36], encourages ethnocentrism more than other identification schemes [37]. Furthermore, these results suggest that intergroup mating itself, not the presence of multiethnic individuals, favors humanitarianism over ethnocentrism.

Three control conditions are included to isolate the key conditions for producing humanitarian cooperation. Each of these controls eliminates one aspect of cooperative intergroup mating in order to identify what is required to produce humanitarian dominance. One control condition eliminates mating, another eliminates intergroup mating, and the third eliminates cooperative mating by having agents mate randomly with one of their Moore neighbors. For simplicity, the Either ID scheme is used for all control conditions. This allows examination of the effects of mating alone, without the presence of mixed-tag offspring. ANOVAs include the Either ID condition with cooperative intergroup mating (Figure 3A). Detailed ANOVA results are presented in Table 6 for strategies and in Table 7 for cooperative behaviors. Plots of means and SEs are presented in Figure 7 for strategies and Figure 8 for behaviors.

Table 6 indicates large and significant main and interaction effects for Strategy. Table 7 shows large, significant main effects for Condition and c/b, as well as for their interaction.

Recall that cooperative intergroup mating evolves humanitarian dominance at low c/b levels (Figure 3a) that does not diminish cooperation compared to ethnocentric dominance at higher c/b levels (Figures 5 and 6). None of the three control conditions yield any humanitarian dominance at any c/b level (Figure 7), confirming that, for humanitarian dominance, there must be mating (7A), that mating must be allowed to occur across group borders (7B), and it must be cooperative (7C).

In the cloning condition, ethnocentrism gives way to selfishness as the c/b ratio increases (7A). When mating is not allowed across group boundaries, selfishness also increases at higher c/b ratios, but does not displace ethnocentrism (7B). When mating is random, rather than cooperative, selfishness prevails at c/b ratios of .6 to 1 (7C). Again, all three components of cooperative intergroup mating must be present for the predominance of a humanitarian pattern of cooperation.

Figure 8 similarly shows that, for all conditions, cooperative behavior diminishes as c/b ratios increase. The main effect of condition is very strong, with lack of intergroup mating producing the most cooperation, followed in order by cooperative intergroup mating, cloning, and random mating. This ordering suggests that cooperative mating by itself, even if restricted to in-group members, fosters cooperation.

Number of intergroup borders is positively correlated with proportion of humanitarian agents and negatively correlated with the proportion of ethnocentric agents in the population. This is true in both the Neither and Either Offspring ID conditions (see Table 8 for details). Also, humanitarians have more intergroup borders than do ethnocentric agents. This is revealed in factorial ANOVAs with independent factors of c/b and strategy (humanitarian vs. ethnocentric). There is a strong main effect of strategy in both the Neither, $F(1, 180) = 136$, $p < .001$, $\eta^2 = .431$, and Either, $F(1, 180) = 926$, $p < .001$, $\eta^2 = .837$, Offspring ID conditions. We omit the Both Offspring ID from these two analyses because intergroup borders tend to disappear in that condition (Figure 4). It is worth noting that humanitarianism also tends to disappear in the Both condition, in favor of ethnocentrism.

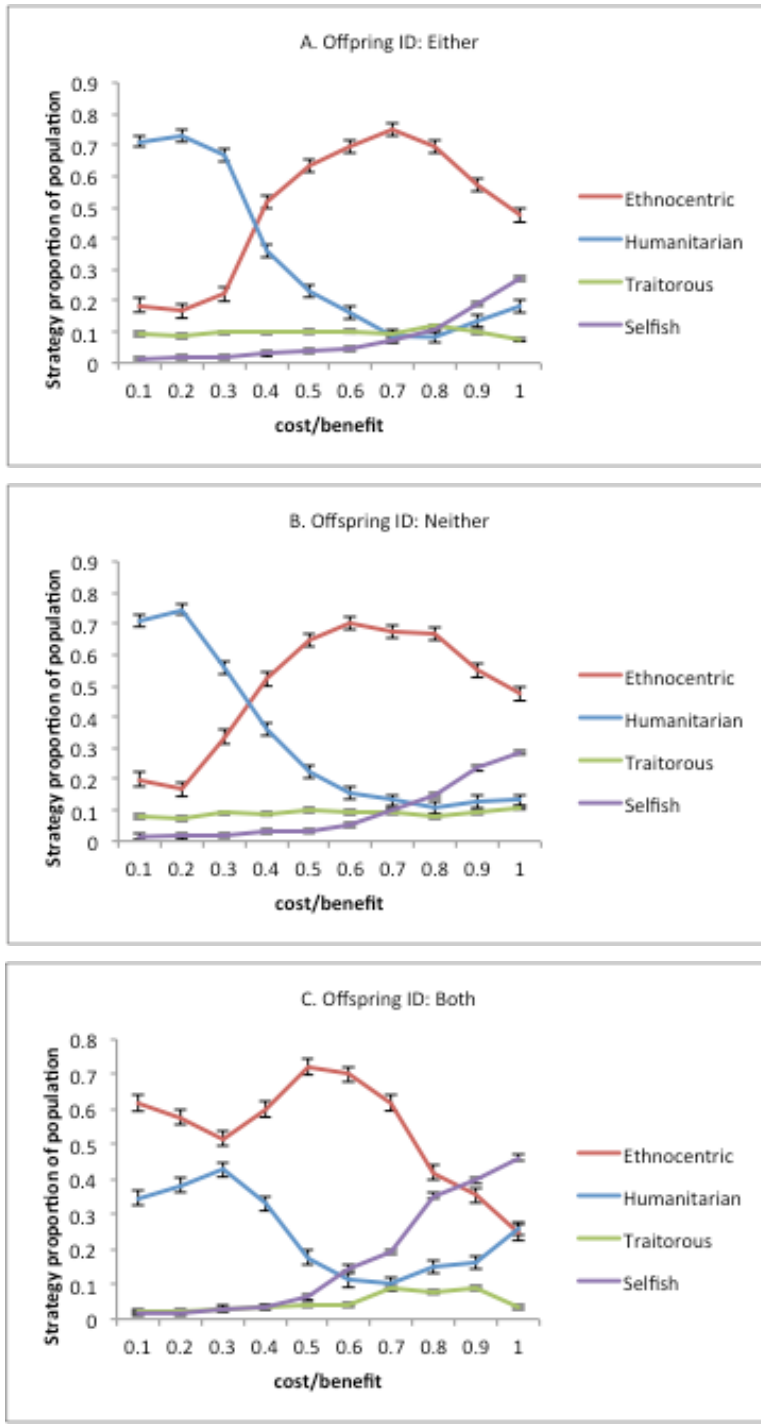

Figure 3. Mean strategy proportions of total population as a function of Offspring ID and c/b. Humanitarian dominance occurs in the Either and Neither conditions at lower c/b levels.

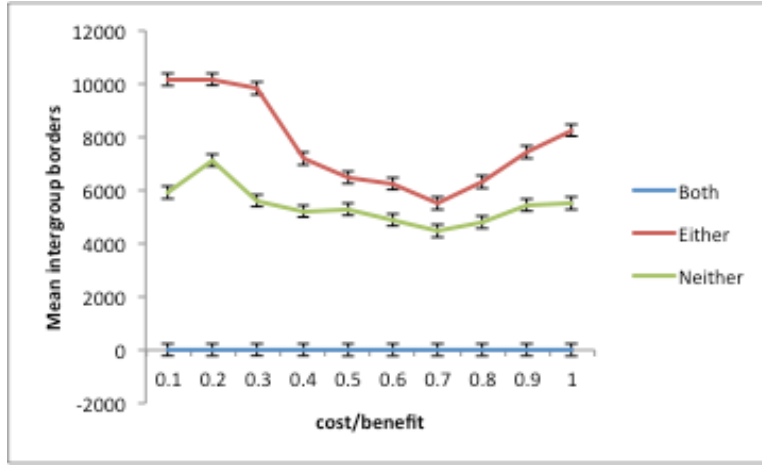

Figure 4. Mean number of intergroup borders as a function of Offspring ID and c/b ratio. Intergroup borders nearly disappear in the Both condition, indicating a homogeneous population.

Table 3. ANOVA Results for Intergroup Borders

| Factor | F | df | $\eta^2$ |
| --- | --- | --- | --- |
| ID | 2812 | 2, 270 | .954 |
| c/b | 34 | 9, 270 | .534 |
| ID x c/b | 15 | 18, 270 | .497 |

Table 4. ANOVA Results for Proportion of Cooperative Behaviors

| Factor | F | df | $\eta^2$ |
| --- | --- | --- | --- |
| ID | 112 | 2, 270 | .453 |
| c/b | 1623 | 9, 270 | .982 |
| ID x c/b | 38 | 18, 270 | .716 |

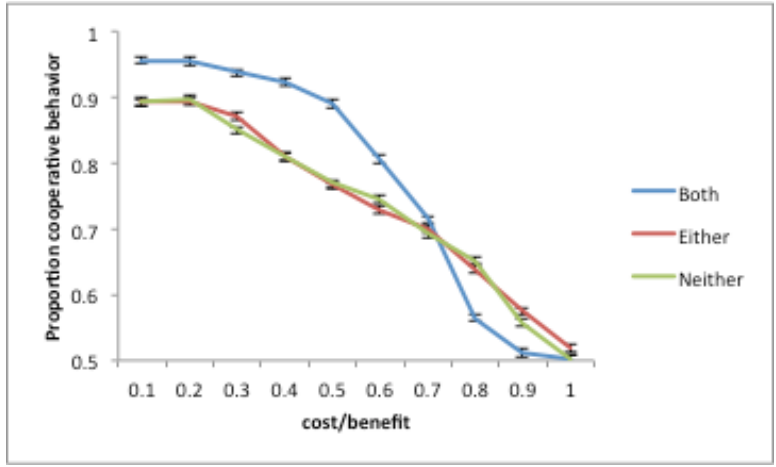

Figure 5. Cooperative behavior as a function of ID and c/b. Cooperation is generally high but decreases as c/b rises.

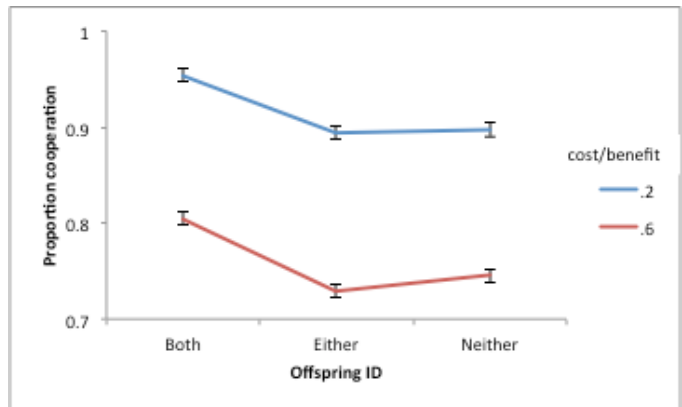

Figure 6. Cooperative behavior as a function of ID re-plotted from Figure 5 at two levels of c/b. Cooperation is higher under humanitarian dominance (at c/b = .2) than under ethnocentric dominance (at c/b = .6).

Table 5. ANOVA Results for Proportion of Cooperative Behaviors at Two c/b Levels

| Factor | F | df | p < | $\eta^2$ |
|---|---|---|---|---|
| ID | 63 | 2, 54 | .001 | .701 |
| c/b | 844 | 1, 54 | .001 | .940 |
| ID x c/b | .87 | 2, 54 | .424 | .031 |

Table 6. ANOVA Effects for Strategy Proportions in Control Conditions

| Factor | F | df | p < | η² |
|---|---|---|---|---|
| Condition | 2.09 | 3, 360 | .101 | 0.017 |
| c/b | .94 | 9, 360 | .487 | 0.023 |
| Condition x c/b | .86 | 27, 360 | .662 | 0.061 |
| Strategy | 5509 | 3, 1080 | .001 | .939 |
| Strategy x Condition | 977 | 9, 1080 | .001 | .891 |
| Strategy x c/b | 262 | 27, 1080 | .001 | .868 |
| Strategy x Condition x c/b | 69 | 81, 1080 | .001 | .837 |

Table 7. ANOVA Results for Proportion of Cooperative Behaviors in Control Conditions

| Factor | F | df | p < | η² |
|---|---|---|---|---|
| Condition | 4686 | 3, 360 | .001 | .975 |
| c/b | 1234 | 9, 360 | .001 | .969 |
| Condition x c/b | 70 | 27, 360 | .001 | .840 |

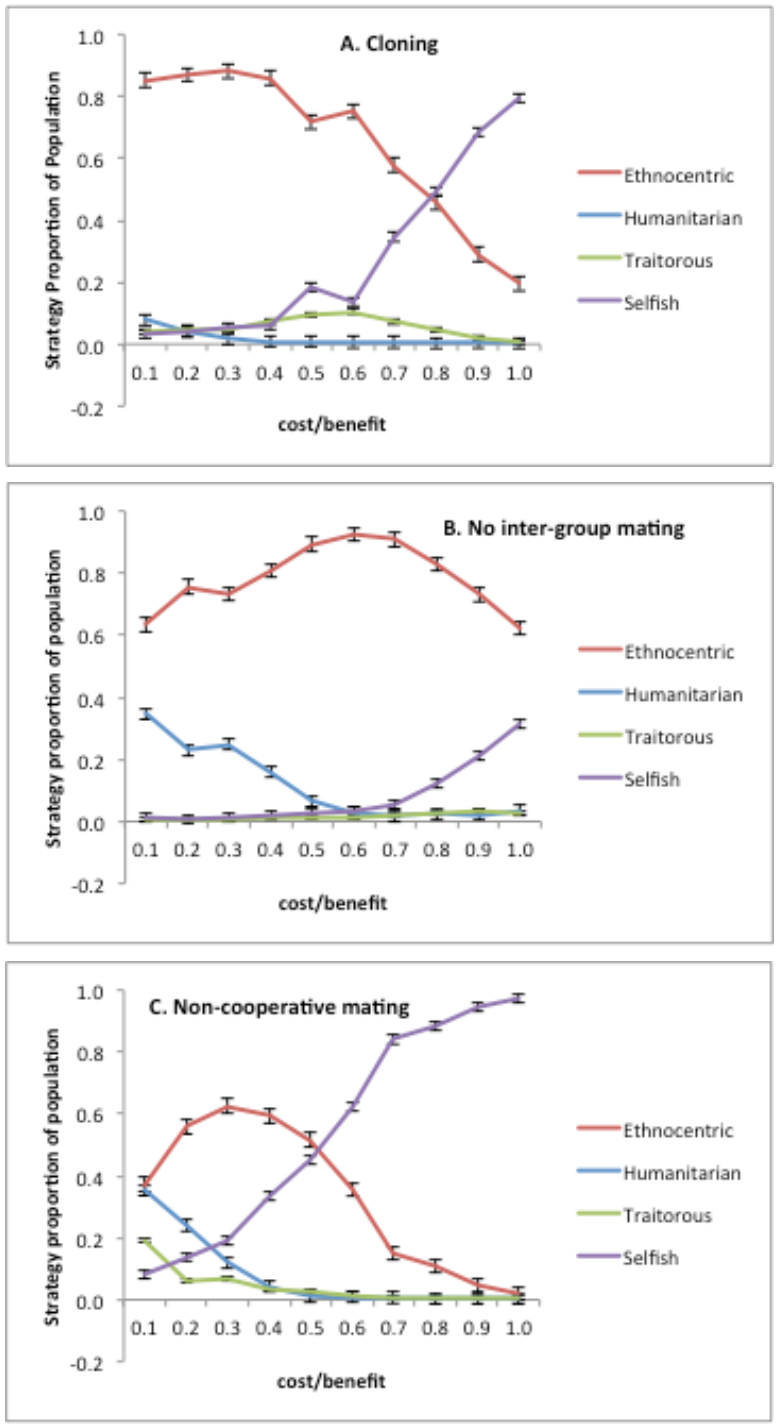

Figure 7. Strategy results for control conditions that eliminate mating (A), intergroup mating (B), or cooperative mating (C) establish that intergroup cooperative mating is critical for humanitarian dominance.

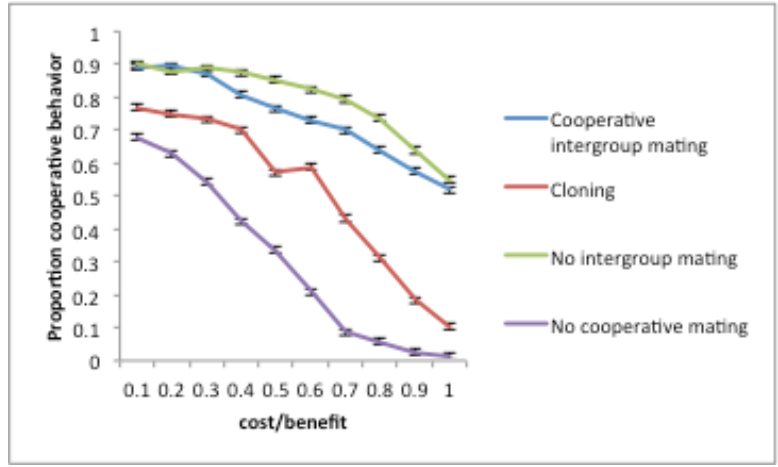

Figure 8. Mean proportions of cooperative behaviors for cooperative intergroup mating and three control conditions that eliminate mating, intergroup mating, or cooperative mating. Cooperative mating enhances general cooperation, whether across or within group boundaries. Cooperation decreases with rising c/b.

Table 8. Pearson Correlations of Number of Intergroup Borders with Proportions of Ethnocentric and Humanitarian Agents in the Population under Cooperative Intergroup Mating

| Offspring ID | Strategy | r | n | df | p < |
|---|---|---|---|---|---|
| Neither | Ethnocentric | -0.524 | 100 | 98 | 0.01 |
| Neither | Humanitarian | 0.511 | 100 | 98 | 0.01 |
| Either | Ethnocentric | -0.173 | 100 | 98 | 0.10 |
| Either | Humanitarian | 0.757 | 100 | 98 | 0.01 |

# Discussion

In our framework, cooperative intergroup mating favors humanitarian strategies under conditions of low cost/benefit ratios and high local diversity. Control conditions show that this effect is driven by the prevalence of cooperative intergroup mating. When offspring are cloned or when intergroup mating is prohibited, ethnocentrism dominates over humanitarianism. When mating is not cooperative, there is either ethnocentric dominance (at low c/b ratios) or selfish dominance (at high c/b ratios). Thus, humanitarian dominance requires cooperative mating across group lines.

Cooperative behavior drops off as c/b ratios increase. Intergroup mating produces the most cooperation, followed in order by cooperative intergroup mating, cloning, and random mating. Thus, cooperative mating by itself fosters cooperation.

Our results highlight the need to incorporate real behaviors into evolutionary simulations. Models that rely exclusively on cooperative-competitive games are appealing in their simplicity, but can miss crucial dynamics at play in the real world [21, 38]. In the present case, traditional models focus on agents' reproductive fitness to calculate the success of a cooperative strategy. However, if mate selection and cooperative strategies are linked, as in our simulations, then cooperation has a direct effect on an agent's reproductive outcome, in addition to its effect on reproductive fitness. In light of this, mate selection can be a potentially powerful force in determining evolutionary outcomes.

Whether reproductive behavior can be accurately modeled using cooperative games remains a point of contention among biologists [39-41], and it seems simplistic to reduce a behavior as complex as mate selection to a decision to cooperate. Nevertheless, reproductive and cooperative behaviors do appear to be tightly coupled. Previous modeling research suggests that social interactions have a strong effect on mating behavior, such that assortative mating may evolve as a consequence of ethnocentric cooperation [42]. Our work extends this finding by establishing a reverse causality, in which mate selection strategies alter the evolutionary course of cooperation. This is analogous to recent findings that several primate species, including humans, are particularly altruistic towards their conspecifics if they have an evolutionary tendency towards cooperative parenting [31]. Although cooperative mating and cooperative parenting are somewhat distinct, they do share the idea that cooperative procreation leads to more general cooperation. Future work could shed additional light on these processes by replicating our simulations on graph structures that are amenable to analytical methods.

Our model delineates the conditions under which intergroup mating, and subsequently humanitarianism, becomes advantageous. First, when the cost/benefit ratio of cooperation decreases, it removes the need for ethnocentric bias to ameliorate the costs of cooperation. In studies of human interethnic couples, marriage can be treated as a mutually beneficial exchange of social and economic resources [41, 43-45]. Unions occur when the cost/benefit ratio is minimal for both parties. Our work predicts that mate choice is more likely to cross ethnic lines if the coupling yields a low cost/benefit ratio. This could occur if, for example, both parties possess high socioeconomic status [43]. Indeed, studies of interethnic marriage find that high-income individuals are more likely to marry across ethnic groups than their low-income counterparts [44]. Thus, improvements in socioeconomic status may facilitate interethnic coupling and humanitarian cooperation strategies. Reductions in the social cost of interethnic marriage, in the form of anti-discrimination legislation, could also prove beneficial.

We find that prevalence of intergroup borders is related positively to humanitarianism and negatively to ethnocentrism. Although direction of causality is difficult to ascertain here, it seems reasonable that causal influence would flow in both directions. Humanitarian parents tend to produce humanitarian offspring, regardless of group tags, thus increasing the number of intergroup borders. Ethnocentric parents tend to produce ethnocentric offspring of the same group tag, thus creating a more homogeneous neighborhood. In the reverse direction, intergroup borders limit mating opportunities for ethnocentric but not for humanitarian agents. It is well known that propinquity is an important aspect of mate selection [45, 46]. In particular, availability plays an important role in the formation of human interethnic unions, with interethnic couples more likely to form in multiethnic neighborhoods [47]. The role of neighborhood diversity in improving inter-group relations has been noted in surveys of humans [48].

We show how evolution can overcome ethnocentrism in favor of the more universal cooperation pattern of humanitarianism. Evolution by itself can achieve humanitarian dominance under a conjunction of conditions: low cost/benefit ratios and the opportunity for cooperative intergroup mating. In this context, evolution-based humanitarianism could be facilitated by social policies that encourage neighborhood diversity and raise the benefit and/or lower the cost of cooperative intergroup mating. Advantages of such evolutionary solutions are that they could be more enduring and less in conflict with competing evolutionary tendencies such as ethnocentrism and selfishness.

A big disadvantage of relying on evolution is that it typically takes a very long time. Societies interested in more immediate solutions may consider environmental interventions based, for example, on encouraging neighborhood diversity and positive contact across group lines, and penalizing discrimination. It seems important for considering such social interventions to not underestimate the apparent evolutionary strength of ethnocentric patterns of cooperation. Effective interventions may have to be strong and persistent to be worthwhile. The rapid demise of intergroup tolerance in the former Yugoslavia after the death of Tito can be cited as a compelling example [49, 50]. All of this implies that future research might fruitfully explore the undoubtedly complex interactions between evolution, social policy, and learning.

In conclusion, ethnocentrism is difficult to minimize because it becomes genetically entrenched due to the fact that it keeps cooperation at a high level, which, in turn, enhances reproductive fitness. Social policies against ethnocentrism may fail to overcome these powerful evolutionary forces favoring ethnocentrism. Our simulations are the first to show that there can be an evolutionary pattern that favors universal over ethnocentric cooperation without lowering the overall level of cooperation. Finally, prolonged social policies could be designed to support that evolutionary solution.

# Acknowledgements

This work is funded by an operating grant to TRS from the Social Sciences and Humanities Research Council of Canada. Peter Helfer, Marcel Montrey, and Artem Kaznatcheev each contributed valuable comments on earlier drafts of this work.